\documentclass[runningheads]{llncs}
\usepackage{graphicx}

\usepackage{multirow}
\usepackage{verbatim}
\usepackage{hyperref}
\usepackage{amsmath,stmaryrd}
\usepackage{algorithm}
\usepackage{algpseudocode}
\usepackage{stix}
\usepackage{hhline}
\usepackage{makecell, multirow}
\usepackage{tikz}
\usepackage{tikzsymbols}

\usepackage{fancyvrb}

\definecolor{mygrey}{rgb}{0.8, 0.8, 0.8}
\usetikzlibrary{arrows.meta}
\usetikzlibrary{calc}
\usetikzlibrary{chains}
\usetikzlibrary{decorations.pathreplacing}
\usetikzlibrary{positioning}
\usetikzlibrary{shapes}
\tikzstyle{arrow} = [draw, -latex']
\tikzset{
    rect/.style={align=center, bottom color=white, top color=white, draw=black, font=\small, minimum size=6mm, minimum width=1mm, node distance=0mm, outer sep=0mm, rectangle},
    empty/.style={align=center, bottom color=white, top color=white, draw=white, font=\small, minimum size=1mm, minimum width=1mm, node distance=0mm, outer sep=0mm, rectangle}
}

\begin{document}
\title{Efficient Algorithms for Modeling SBoxes Using MILP }
%
%

\author{Debranjan Pal \and
Vishal Pankaj Chandratreya \and\\
Dipanwita Roy Chowdhury}
\authorrunning{Debranjan Pal et al.}
%
\institute{Crypto Research Lab, IIT Kharagpur, India \\ 
\email{debranjan.crl@gmail.com, vpaijc@kgpian.iitkgp.ac.in, \\drc@cse.iitkgp.ac.in}}

\maketitle              
\begin{abstract}
Mixed Integer Linear Programming (MILP) is a well-known approach for the cryptanalysis of a symmetric cipher. A number of MILP-based security analyses have been reported for non-linear (SBoxes) and linear layers. Researchers proposed word- and bit-wise SBox modeling techniques using a set of inequalities which helps in searching differential trails for a cipher. 

In this paper, we propose two new techniques to reduce the number of inequalities to represent the valid differential transitions for SBoxes. Our first technique chooses the best greedy solution with a random tiebreaker and achieves improved results for the 4-bit SBoxes of MIBS, LBlock, and Serpent over the existing results of Sun et al.~\cite{Sun_asiacrypt}.
Subset addition, our second approach, is an improvement over the algorithm proposed by Boura and Coggia. Subset addition technique is faster than Boura and Coggia~\cite{BouraC} and also improves the count of inequalities. Our algorithm emulates the existing results for the 4-bit SBoxes of Minalpher, LBlock, Serpent, Prince, and Rectangle. The subset addition method also works for 5-bit and 6-bit SBoxes. We improve the boundary of minimum number inequalities from the existing results for 5-bit SBoxes of ASCON and SC2000. 
Application of subset addition technique for 6-bit SBoxes of APN, FIDES, and SC2000 enhances the existing results. By applying multithreading, we reduced the execution time needed to find the minimum inequality set over the existing techniques.

\keywords{Mixed Integer Linear Programming  \and Symmetric Key \and Block Cipher \and Active SBox \and Differential Cryptanalysis \and Convex Hull}
\end{abstract}
\section{Introduction}
Differential cryptanalysis~\cite{DifferentialCrypt_BihamShamirDCDES} and linear cryptanalysis~\cite{LinearCrypt_MatsuiY92} are the two most valuable methods in the cryptanalysis domain of symmetric-key cryptography.
Differential cryptanalysis for block ciphers demonstrates the mapping of input differences in the plaintext to output differences in the ciphertext. The probabilistic linear relationships between the plaintext, ciphertext and key are expressed by linear cryptanalysis, on the other hand. For developing a distinguisher or identifying a key-recovery attack, we leverage the feature that an ideal cipher behaves differently from a random cipher for differential or linear cryptanalysis. 

For new ciphers, enforcing resistance against the linear and differential cryptanalysis is a common design criterion during the analysis of the cipher. The wide trail design technique for block ciphers leads to prove the security against linear and differential cryptanalysis.  Finding the minimum number of active SBoxes within the cipher is a useful way to assess the security of the scheme against differential attacks. Calculating the minimal amount of active SBoxes for block ciphers has received a lot of attention. But the time and effort by humans as well as by programs required to apply those techniques is huge. Here, MILP can be used to solve these issues. MILP is an optimization problem that restricts some of its variables to integral value. MILP offers a powerful and adaptable approach for handling large, significant and complicated problems. The attacker uses linear inequalities to define potential differential propagation patterns in a round function for differential search using MILP. The attacker then executes an MILP solver (in parallel), which yields the minimal number of active SBoxes for the specified propagation patterns. After generating a lower bound for the quantity of active SBoxes in differential and linear cryptanalysis, one may calculate an upper bound for the likelihood of the best characteristic by utilising the maximum differential probability (MDP) of the SBoxes. The probability of the best characteristic can be used to calculate the differential probability with accuracy by summing up all matching characteristics probabilities. 

Typically, a cryptographic component's (like SBox) propagation characteristic can be expressed using Boolean functions. To model a Boolean function using the MILP method, a set of inequalities must be computed. Then some solutions to these inequalities also calculated which precisely corresponds to support the boolean function. In order to calculate a solution model of any boolean function, we must resolve two issues. To efficiently model a boolean function a set of inequalities to be created. Selecting a minimal set of possible inequalities which perfectly model the boolean function, if the first issue is resolved. To solve the these problems researchers proposes different techniques~\cite{Mouha,Sun_asiacrypt,sun_towards,Sun_SbP,BouraC,Sasaki_MILP_ALG,Superball_Li_Sun_tosc}.

MILP models are frequently used in cryptanalysis, it is crucial to increase the effectiveness of MILP model solutions. Which kind of comprehensive model, though, would be the most effective is still a mystery. Different kinds of proposed models should be built and thoroughly investigated in order to fix this issue. Since different models are built from a varieties of possible inequalities, the first move towards achieving this goal is to generate a variety of inequalities.

To get optimized MILP model researchers observe reduction in number of inequalities by at least one unit is itself important. Because in a larger view, for a full round cipher, the total number of inequalities impacts a lot with respect to the computing resources and the timing requirements  for the MILP solver.

\subsection{Related Work}
 Mouha et al.~\cite{Mouha} first described the problem of finding the smallest number of active SBoxes that can be modelled using MILP for assessing word-oriented ciphers.
There are other ciphers that are not word-oriented, such as PRESENT, which applies a 4-bit SBox and then switches four bits from one SBox to four separate SBoxes using a bit-permutation. 

Sun et al.~\cite{Sun_asiacrypt} established a way to simulate all possible differential propagations bit by bit even for the SBox in order to apply MILP to such a structure. 
They used MILP-based techniques to assess a block cipher's security for related-key differential attacks. The methods are mainly applied for searching single-key or related-key differential~\cite{RelatedKetDiff_Biham93} characteristics on PRESENT80, LBlock, SIMON48, DESL and PRESENT128. Sun et al.~\cite{Sun_asiacrypt} report different methods to model differential characteristics of an SBox with linear inequalities. 
In the first technique, inequalities are produced based on some conditional differential characteristics of an SBox. Another method involves extracting the inequalities from the H-representation of the convex hull for all possible differential patterns of the SBox.
To choose a certain number of inequalities from the convex hull, they devise a greedy algorithm for the second technique. They suggest an automated approach for assessing the security of bit-oriented block ciphers for differential attack. They propose many ways for getting tighter security limits using these inequalities along with the MILP methodology.

Sun et al.~\cite{sun_towards} look into the differential and linear behaviour of a variety of block ciphers using mixed-integer linear programming (MILP). They point out that a modest set of linear inequalities can precisely characterize the differential behaviour of every SBox. For a variety of ciphers, Sun et al.~\cite{sun_towards} build MILP models whose feasible zones are exactly the collections of all legitimate differential and linear properties. Any subset of $\{0,1\}^n\subset\mathbb{R}^n$ has an accurate linear-inequality description, according to them. They provide a technique that may be used to determine all differential and linear properties with certain specified features by converting the heuristic approach of Sun et al.~\cite{Sun_asiacrypt} for finding differential and linear characteristics into an accurate one based on these MILP models. 

Mouha's~\cite{Mouha} technique is not suitable for SPN ciphers which contain diffusion layers with bitwise permutations, called S-bP structures. The problem occurs because of avoiding the diffusion effect calculated simultaneously by the non-linear substitution layers and bitwise permutation layers. Also the MILP constraints provided by Mouha are not sufficient for modeling the differential propagation of a linear diffusion layer derived from almost-MDS or non-MDS matrix. To automatically determine a lower constraint on the number of active SBoxes for block ciphers with S-bP structures, Sun et al.~\cite{Sun_SbP} expanded the method of Mouha et al.~\cite{Mouha} and proposed a new strategy based on mixed-integer linear programming (MILP). They successfully applied the technique to PRESENT-80. 

In order to automatically look for differential and linear characteristics for ARX ciphers, Kai Fu et al.~\cite{KaiFui_SPECK} built a MILP model. By assuming independent inputs to the modular addition and independent rounds, they applied the differential and linear property of modular addition.
They search for the differential properties and linear approximations of the Speck cipher using the new MILP model. Their identified differential characteristics for Speck64, Speck96, and Speck128 are prolonged for one, three, and five rounds, respectively, in comparison to the prior best differential characteristics for them, and the differential characteristic for Speck48 has a greater likelihood. Cui et al.~\cite{Cui} provide a novel automatic method to search impossible differential trails for ciphers containing SBoxes after taking into account the differential and linear features of non-linear components, like SBoxes themselves.
They expand the tool's capabilities to include modulo addition and use it with ARX ciphers.
For HIGHT, SHACAL-2, LEA, and LBlock, the tool enhances the best outcomes currently available. 
A new SBox modeling that can handle the likelihood of differential characteristics and reflect a condensed form of the DDT of big SBoxes was presented by Ahmed Abdelkhalek et al.~\cite{Abdelkhalek_Large_SBOX}.
They increased the number of rounds needed to resist simple differential distinguishers by one round after evaluating the upper bound on SKINNY-128's differential features.
For two AES-round based constructions, the upper bound on differential features are examined. 
 
\subsection{Our Contribution} 
In this paper we introduce two of MILP-based solutions for valid differential trail propagation through the non-linear layers, that is, SBoxes. The new approaches reduce the number of inequalities for modeling 4-bit SBoxes in comparison to the earlier algorithms~\cite{BouraC,Sun_asiacrypt,Sasaki_MILP_ALG,DBLP:journals/iacr/Udovenko21}. Our techniques help cryptographic designers for providing a bound on finding minimum number of SBoxes and thus ensure resistance against differential cryptanalysis attacks.
\begin{itemize}
    \item \textbf{Greedy random-tiebreaker algorithm} We propose a new algorithm by randomly choosing from the result of greedy algorithm. Our technique improves the boundary for minimum number of inequalities for 4-bit SBoxes of MIBS, LBlock and Serpent over the existing greedy algorithm~\cite{Sun_asiacrypt}.
    \item \textbf{Subset addition approach} A subset-addition-based algorithm is proposed by generating new inequalities from the results of H-representation of the convex hull. We add $k$-subset inequalities to generate new inequality, which removes more impossible propagation. Then replace some subset of old inequalities by the newer one, which results in an improvement over the existing algorithms for 4-bit SBoxes of  Minalpher, LBlock, Serpent, Prince and Rectangle. The subset addition algorithm also works for 5- and 6-bit SBoxes. We also improve the boundary of inequalities for 5-bit SBoxes of ASCON, and SC2000. For 6-bit SBoxes of APN and SC2000, we reduce the number of inequalities from the existing results.

    We also improve the time for finding the minimum set of inequalities 
    over Boura and Coggia's~\cite{BouraC} approach by a significant percentage. 
    
\end{itemize}

\subsection{Organization of the paper}
The organization of the paper is as follows. Section~\ref{Background} explains the background of our work. We describe the greedy random tiebreaker algorithm and its results in Section~\ref{GreedyRandomTiebreaker}. In Section~\ref{SubsetAddition}, we present the subset addition approach and the corresponding implementation process with the results. Section~\ref{Conclusion} concludes the paper.

\section{Background}\label{Background}
In this section we describe the earlier used methods and algorithms for modeling SBoxes using inequalities.

\subsection{Representation of SBoxes using inequalities}
An SBox $S$ can be represented as $S:\mathbb{F}_2^n \rightarrow \mathbb{F}_2^n$. We can symbolize any operation on an SBox as $x \rightarrow y$ with $x, y \in \mathbb{F}_2^n$.
Let $(x_0,\ldots, x_{n-1}, y_0,\ldots, y_{n-1}) \in \mathbb{R}^{2n}$ be a $2n$-dimensional vector, where $\mathbb{R}$ is the real number field, and for an SBox the input-output differential pattern is denoted using a point $(x_0,\ldots, x_{n-1}, y_0,\ldots, y_{n-1})$.



\subsubsection{H-representation of the convex hull}

The convex hull of a set $P$ of distance points in $\mathbb{R}^n$ is the smallest convex set that contains $P$. We compute the H-representation of the convex hull of all possible input-output differential patterns of the SBox by calculating the DDT. Applying SageMath~\cite{sagemath} on the DDT, we compute the H-representation. 
From H-representation we get $w$ linear inequalities, which can be written as

\begin{align*}
    A
    \begin{bmatrix}
        x_0\\
        \vdots\\
        x_{n-1}\\
        y_0\\
        \vdots\\
        y_{n-1}
    \end{bmatrix}
    \leq b
\end{align*}
where $A$ is a $w \times 2n$ matrix. Here, $A$ and $b$ contain only integer values.
Every linear inequality also invalidates some points, which are associated with some impossible differential propagations. 
The H-representation also contains redundant inequalities associated with the MILP-based differential trail search; the reason is that the feasible points are constricted to $\{0, 1\}^{2n}$, not $\mathbb{R}^{2n}$ As a result, a lot of extra linear inequalities force the MILP solver to work slower. To eliminate those redundant inequalities, researchers apply different techniques.

\subsubsection{Conditional differential characteristics modeling}
The logical condition that $(x_0,\ldots,x_{m-1})=(\delta_0,\ldots,\delta_{m-1}) \in \{0,1\}^m \subseteq \mathbb{Z}^m$ implies $y=\delta \in \{0,1\}\subseteq \mathbb{Z}$ can be modeled by using the following linear inequality,
        \begin{equation}
                \sum_{i=0}^{m-1}(-1)^{\delta_i}x_i+ (-1)^{\delta+1}y-\delta+\sum_{i=0}^{m-1}\delta_i \geq 0
        \end{equation}
Let $(x_0, x_1, x_2, x_3)$ and $(y_0, y_1, y_2, y_3)$ be MILP variables for the input and output differences of a 4-bit SBox. Let $(1010) \rightarrow (0111)$ be an impossible propagation in the DDT corresponding to the SBox. That is, the input difference $(1010)$ is not propagating to $(0111)$. By equation 1, the linear inequality which eliminates the impossible point $(1001, 1101)$ is, $-x_0 + x_1 - x_2 + x_3 + y_0 - y_1 - y_2 - y_3 + 4 \geq 0$. 
Corresponding to an SBox, if in a DDT there occur $n$ impossible paths, then at most $n$ linear inequalities are needed to model the DDT correctly. But one can reduce the value of $n$ by merging one or more available inequalities and therefore generating new inequalities.
For example if we consider two impossible propagations $(1010) \rightarrow (0111)$ and $(1010) \rightarrow (0110)$, then the linear inequality:
$-x_0 + x_1 - x_2 + x_3 + y_0 - y_1 - y_2 + y_3 \geq -3$
eliminates both the impossible points together. 

\subsection*{Choosing the best inequalities from the convex hull}
Applying an MILP model, generating a feasible solution is not guaranteed to be a legitimate differential path. We want to reduce the number of active SBoxes throughout a greater region and the optimal value achieved should be less than or equal to the actual minimum number of active SBoxes.
Hence, we will look for any linear inequality that may be used to chop out a portion of the MILP model while maintaining the region of valid differential characteristics.  Researchers proposed different algorithms (see Appendix~\ref{Appendix_Algorithms}) for reducing number of inequalities for an SBox representation.
\subsubsection*{Greedy Algorithm Based Modeling~\cite{Sun_asiacrypt}}
The discrete points in the H-Representation (convex hull) generate a huge number of inequalities.
A good approach is to filter the best valid inequalities which maximize the number of removed impossible differential patterns from the feasible region of the convex hull. Algorithm ~\ref{alg:greedy_sun} explains the greedy approach proposed by Sun et al.

\subsubsection*{Modeling by selecting random set of inequalities~\cite{BouraC}}
A larger set of new inequalities (generated by randomly adding $k$ inequalities) are worthless as most probably they will satisfy the whole space $\{0,1\}^m$.
If $k$-hyperplanes of the H-representation share a vertex on the cube $\{0,1\}^m$, then summing corresponding inequalities probably generates a new inequality $Q_{new}$.
But the hyperplane corresponding $Q_{new}$ should intersect with the cube by at least one point.
In that case, $Q_{new}$ possibly invalidates a different set of impossible points from H-representation than the older inequalities. Algorithm~\ref{alg:boura_select_any} briefs the whole procedure.

\begin{figure}[htbp]
    \centering
    \begin{tikzpicture}
        \node(start)[rect]{$\{1,2\},\{1,2,3\},\{2,3,5\},\{4,5\},\{6\},\{6,7\}$};

        \node(g1)[rect, below left=8mm and 5mm of start.south]{$\{\textcolor{mygrey}{1},\textcolor{mygrey}{2}\},\{\textcolor{mygrey}{2},\textcolor{mygrey}{3},5\},\{4,5\},\{6\},\{6,7\}$};
        \node(g2)[rect, below=8mm of g1]{$\{\textcolor{mygrey}{1},\textcolor{mygrey}{2}\},\{\textcolor{mygrey}{2},\textcolor{mygrey}{3},\textcolor{mygrey}{5}\},\{6\},\{6,7\}$};
        \node(g3)[rect, below=8mm of g2]{$\{\textcolor{mygrey}{1},\textcolor{mygrey}{2}\},\{\textcolor{mygrey}{2},\textcolor{mygrey}{3},\textcolor{mygrey}{5}\},\{\textcolor{mygrey}{6}\}$};
        \path[arrow](start.south)--node[midway, left, align=center]{Choose $\{1,2,3\}$\hspace{5mm}}(g1.north);
        \path[arrow](g1)--node[midway, left, align=center]{Choose $\{4,5\}$}(g2);
        \path[arrow](g2)--node[midway, left, align=center]{Choose $\{6,7\}$}(g3);

        \node(h1)[rect, below right=8mm and 5mm of start.south]{$\{1,\textcolor{mygrey}{2}\},\{1,\textcolor{mygrey}{2},\textcolor{mygrey}{3}\},\{4,\textcolor{mygrey}{5}\},\{6\},\{6,7\}$};
        \node(h2)[rect, below=8mm of h1]{$\{1,\textcolor{mygrey}{2}\},\{1,\textcolor{mygrey}{2},\textcolor{mygrey}{3}\},\{4,\textcolor{mygrey}{5}\},\{\textcolor{mygrey}{6}\}$};
        \node(h3)[rect, below=8mm of h2]{$\{1,\textcolor{mygrey}{2}\},\{1,\textcolor{mygrey}{2},\textcolor{mygrey}{3}\},\{\textcolor{mygrey}{6}\}$};
        \node(h4)[rect, below=8mm of h3]{$\{\textcolor{mygrey}{1},\textcolor{mygrey}{2},\textcolor{mygrey}{3}\},\{\textcolor{mygrey}{6}\}$};
        \path[arrow](start.south)--node[midway, right, align=center]{\hspace{5mm}Choose $\{2,3,5\}$}(h1.north);
        \path[arrow](h1)--node[midway, right, align=center]{Choose $\{6,7\}$}(h2);
        \path[arrow](h2)--node[midway, right, align=center]{Choose $\{4,5\}$}(h3);
        \path[arrow](h3)--node[midway, right, align=center]{Choose $\{1,2\}$}(h4);
    \end{tikzpicture}
    \caption{Variation of the output of the greedy algorithm because of a random tiebreaker}
\end{figure}
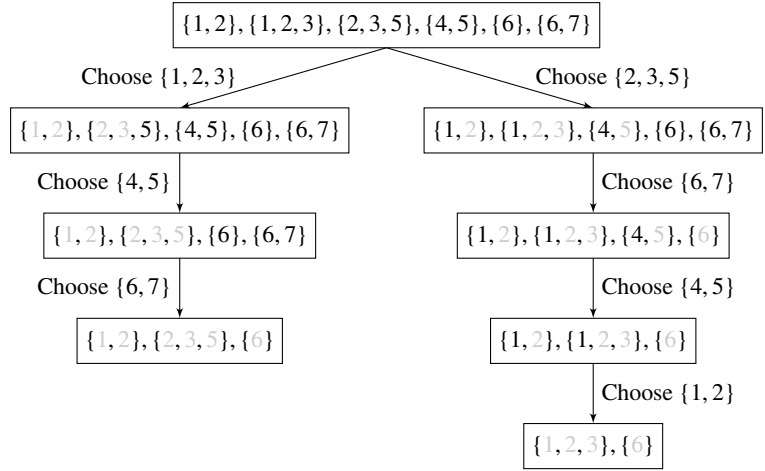

 \begin{algorithm}[htbp]
    Input:\\ 
    $H_{Rep}$: Inequalities in the H-Representation of the convex hull of an SBox.\\
    $I_D$: The set of all impossible differential paths of an SBox 
    \\
    Output:\\
    $I_{SR}$: Set of n-best inequalities that generates more stricter feasible region after maximizing the removed impossible differential paths. 
	\begin{algorithmic}[1]
   \State $I_M \gets \phi, I_{SR} \gets \phi$
   \While {$I_D \ne \phi$}
    \State $I_M \gets $The inequalities in $H_{Rep}$ which maximizes the number of removed impossible differential paths from $I_D$.
    \State $I_D \gets$ $I_D-\{\text{Removed impossible differential paths using $I_M$\}}$
    \If {$Degree(I_M) > 1$}\Comment{Returns the number of elements in a set}
    \State $Rand_{I} \leftarrow \textit{ChooseRandomInequality}(I_M)$\Comment{Chooses randomly an element from a set}
    \EndIf
    \State $H_{Rep} \gets H_{Rep} - Rand_I$
    \State $I_{SR} \gets I_{SR} ~\displaystyle{\cup}~ Rand_I$
    \EndWhile
    \State \textbf{return $I_{SR}$}
    \end{algorithmic}
    \caption{Randomly select inequalities from greedy set (Greedy Random-Tiebreaker)}
	\label{alg:greedy_random}
	\end{algorithm}

\subsubsection{MILP-based reduction algorithm}
Sasaki and Todo\cite{Sasaki_MILP_ALG} propose a technique which generates a minimization problem, which they solve using a standard MILP solver to get the reduced set of inequalities. First they find all the impossible differential points from DDT of an SBox and generate impossible patterns applying those points. Next, for a given impossible pattern, check which subset of inequalities invalidates the pattern. Now, they form a constraint: every impossible pattern should be removed from the possible region by at least one inequality. They form an MILP problem which minimizes the total set of inequalities applying the constraints and solve it to get the minimized set of inequalities. Algorithm~\ref{alg:sasaki_MILP}
describes the whole process in a step-wise manner.

\begin{algorithm}[htbp]
    Input:\\ 
    $H_{Rep}$: Inequalities in the H-Representation of the convex hull of an SBox.\\
    $I_P$: The set of all possible differential paths of an SBox 
    \\
    $I_D$: The set of all impossible differential paths of an SBox 
    \\
    Output:\\
    $I_{SR}$: Set of n-best inequalities that generates more stricter feasible region after maximizing the removed impossible differential paths. 
	\begin{algorithmic}[1]
   \State $I_{SR} \gets H_{Rep}$
   \For {$p \in I_P $}
    \State $I_H \leftarrow$ all hyperplanes in $H_{Rep}$ which the point $p$ lies on
    
    \For {$\{h_1, h_2,\ldots, h_k\} \in P_{Set}(I_H)$}
    \State $h \leftarrow h_1 + h_2 + \ldots + h_k$
    \If{$h$ is a good hyperplane (If satisfies condition of Type 1 or Type 2)}
    \State $I_{SR} \leftarrow I_{SR} \cup \{h\}$
    \EndIf
    \EndFor
    \EndFor
    \State  \textbf{return} the smallest subset of $I_{SR}$ removing all paths in $I_D$.
    \end{algorithmic}
    \caption{Generates a minimal subset of inequalities eliminating all impossible differential paths (Subset Addition)}
	\label{alg:subset_addition}
	\end{algorithm}

\begin{figure}[htbp]
    \centering
    \includegraphics[height=6.9cm, keepaspectratio]{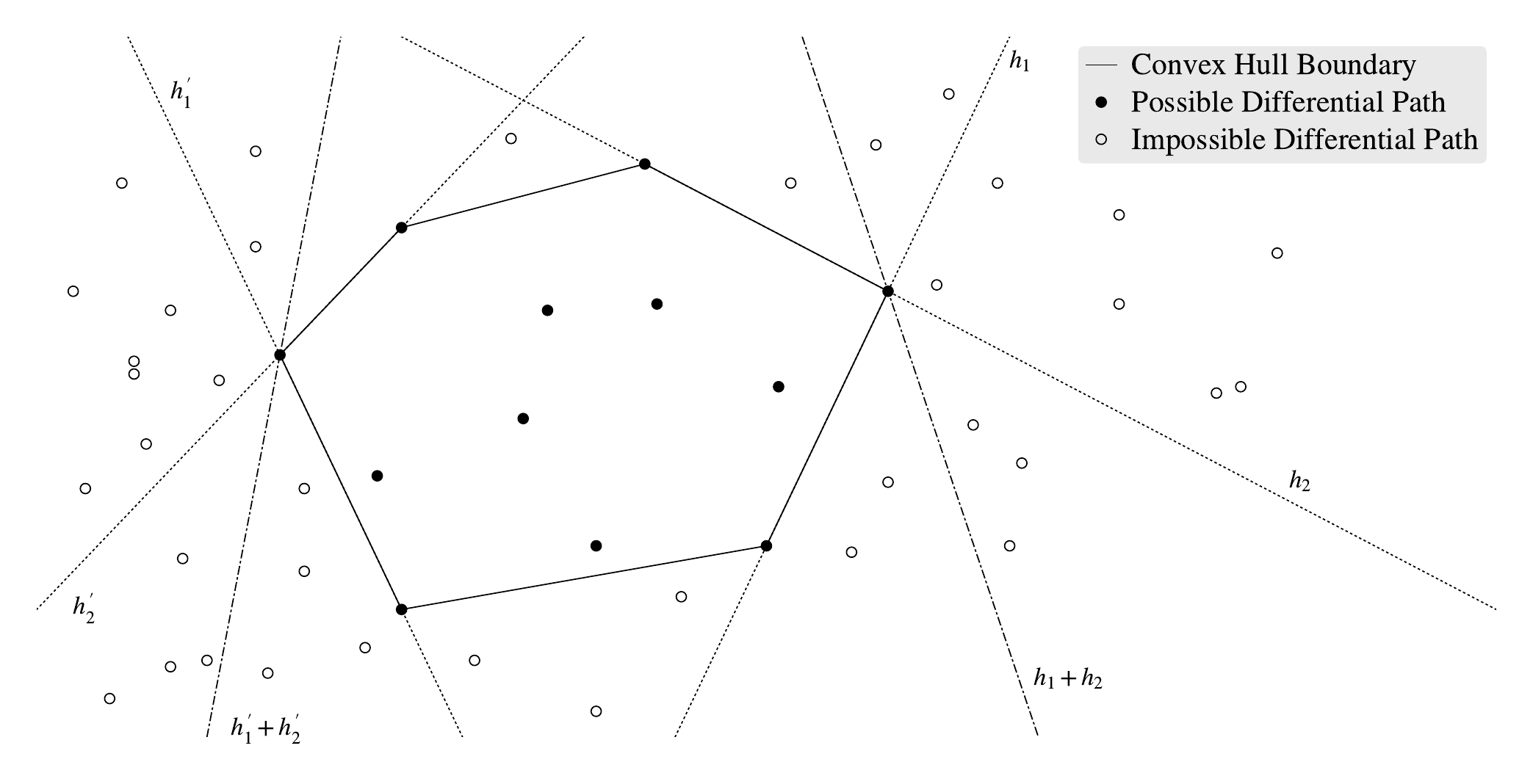}
    \caption{Separating Type 1 and Type 2 inequalities in convex hull}
    \label{fig:m_milp_diagram}
\end{figure}

\section{Filtering Inequalities by Greedy Random-Tiebreaker}\label{GreedyRandomTiebreaker}
We use the original greedy algorithm proposed by S Sun et al.~\cite{Sun_asiacrypt}. Our technique is similar to the original greedy algorithm, except that when multiple inequalities have the same rank, we choose one of them randomly.
We claim that a random tiebreaker could result in a different number of inequalities across multiple runs, and prove it with an instance of the set cover problem (which is homomorphic to the problem of minimising the MILP model of an SBox). Consider the set
$S = \{1, 2, 3, 4, 5, 6, 7\}$ and its subsets $\{\{1, 2\}, \{1, 2, 3\}, \{2, 3, 5\}, \{4, 5\}, \{6\}, \{6, 7\}\}$, which are to be used to find a cover of $S$. Two possible greedy approaches yield covers of sizes three, $\{\{1, 2, 3\}, \{4, 5\}, \{6, 7\}\}$ and four $\{\{2, 3, 5\}, \{4, 5\}, \{6, 7\}, \{1, 2\}\}$.

The overall technique is described in Algorithm~\ref{alg:greedy_random}. Let $H_{Rep}$ be the set of inequalities that is generated from the SageMath method $inequality\_generator()$ as the H-representation of the convex hull of an SBox. Assume $I_D$ be the set of all impossible differential points selected from difference distribution table (DDT) of an SBox. Let $I_M$ stores the hyperplanes in $H_{Rep}$ removing greatest number of paths from $I_D$. If $I_M$ have more than one elements, then we are choosing an inequality randomly from $I_M$ by applying  \textit{ChooseRandomInequality} method and collecting in $I_{SR}$. The best set of inequalities saved to $I_{SR}$. We execute the overall process multiple times to get the  best set of inequalities.

\begin{table}
    \centering
    \caption{Minimum number of Inequalities for 4-bit SBoxes (Random Greedy-Tiebreaker)}
    \begin{tabular}{ |c|c|c|c| }
        \hline
        Cipher&\thead{SageMath~\cite{sagemath}}&\thead{Sun et al.~\cite{Sun_asiacrypt}}&\thead{Random Greedy (Our Approach)}\\
        \hline
        GIFT&237&-&22\\
        KLEIN&311&22&22\\
        Lilliput&324&-&26\\
        MIBS&378&27&\bf24\\
        Midori S0&239&-&25\\
        Midori S1&367&-&24\\
        Minalpher&338&-&25\\
        Piccolo&202&23&24\\
        PRESENT&327&22&22\\
        PRIDE&194&-&22\\
        PRINCE&300&26&26\\
        RECTANGLE&267&-&23\\
        SKINNY&202&-&24\\
        TWINE&324&23&25\\
        \hline
        LBlock S0&205&28&\bf25\\
        LBlock S1&205&27&\bf25\\
        LBlock S2&205&27&\bf25\\
        LBlock S3&205&27&\bf26\\
        LBlock S4&205&28&\bf25\\
        LBlock S5&205&27&\bf25\\
        LBlock S6&205&27&\bf26\\
        LBlock S7&205&27&\bf25\\
        LBlock S8&205&28&\bf26\\
        LBlock S9&205&27&\bf25\\
        \hline
        Serpent S0&410&23&24\\
        Serpent S1&409&24&25\\
        Serpent S2&408&25&25\\
        Serpent S3&396&31&\bf23\\
        Serpent S4&328&26&\bf24\\
        Serpent S5&336&25&\bf23\\
        Serpent S6&382&22&\bf21\\
        Serpent S7&470&30&\bf21\\
        Serpent S8&364&-&25\\
        Serpent S9&357&-&24\\
        Serpent S10&369&-&27\\
        Serpent S11&399&-&21\\
        Serpent S12&368&-&24\\
        Serpent S13&368&-&24\\
        Serpent S14&368&-&25\\
        Serpent S15&368&-&23\\
        Serpent S16&365&-&25\\
        Serpent S17&393&-&31\\
        Serpent S18&368&-&27\\
        Serpent S19&398&-&23\\
        Serpent S20&351&-&24\\
        Serpent S21&447&-&25\\
        Serpent S22&405&-&25\\
        Serpent S23&328&-&24\\
        Serpent S24&357&-&24\\
        Serpent S25&366&-&22\\
        Serpent S26&368&-&23\\
        Serpent S27&523&-&24\\
        Serpent S28&278&-&23\\
        Serpent S29&394&-&24\\
        Serpent S30&394&-&23\\
        Serpent S31&357&-&27\\
        \hline
    \end{tabular}
    \label{tab:min_ineq_random_greedy}
\end{table}

\subsection{Implementations and results}
We implemented Algorithm~\ref{alg:greedy_random} in SageMath (on a desktop computer with an Intel Core i5-6500 4C/4T CPU running Manjaro Linux Sikaris 22.0.0 Xfce Edition 64-bit) with a flag to randomise
the list of inequalities.

\subsubsection*{Applications on 4-bit SBoxes}
We run our algorithm greedy random-tiebreaker on a large set 4-bit SBoxes used in ciphers. Now we describe the comparison of our results with existing best known result against the minimum number of inequalities. The first block of Table~\ref{tab:min_ineq_random_greedy} presents the results for 4-bit SBoxes for 14 different ciphers. We get mixed results, for instance in case of MIBS~\cite{DBLP:mibs} we get 24 inequalities which is lesser in 3 numbers that of Sun et al.~\cite{Sun_asiacrypt}; for Prince~\cite{DBLP:prince} the result is same but for other four ciphers it gives comparable results. 
The third block of Table~\ref{tab:min_ineq_random_greedy} shows the result for all the SBoxes of Serpent~\cite{DBLP:serpent}. Among the existing available results of eight SBoxes for four SBoxes $(S3,S4,S5,S7)$ we are gaining, for S2 the results is same and for the remaining three $(S0, S1, S6)$ we are loosing at most in two numbers. In Table~\ref{tab:min_ineq_random_greedy} second block, for LBlock~\cite{DBLP:lblock} the results are better over Sun et al~\cite{Sun_asiacrypt} for all the SBoxes.
For MIBS~\cite{DBLP:mibs}, the reduced 24 inequalities are provided in Appendix~\ref{Appendix_SampleInequalities}.

\begin{algorithm}[htbp]
 \textbf{Input:}The SBox and a positive integer $k$\\
 \textbf{Output:} Minimized set of inequalities $I_{SR}$
   \begin{algorithmic}[1]
   \State \textit{ddt} $\leftarrow$ \textit{DDT(SBox)}\Comment{Create and returns the difference distribution table}
       \State \textit{impPoints} $\leftarrow$ \textit{GetImpossiblePaths (ddt)}\Comment{Returns the impossible transitions (points)}
   \State \textit{validPoints} $\leftarrow$ \textit{GetValidPaths (ddt)} \Comment{Returns the valid transitions (points)}
   \State I $\leftarrow$ \textit{inequality\_generator (validPoints)}\Comment{Returns the inequalities}
   
   \State \textit{Path} $\leftarrow \phi$\Comment{Each element of Path consists of two parts, an inequality and a point}
   \For {iq in I}
   \State \textit{point\_iq} $\leftarrow \phi$
   \For {\textit{point} in \textit{impPoints}}
   \If { \textit{Evaluate(iq, point)} < 0}
   \State  \textit{point\_iq} $\leftarrow$ \textit{point\_iq} $\cup$ \textit{point}
   \EndIf
   \EndFor
   \State \textit{Path} $\leftarrow$ \textit{Path} $\cup$ \{iq, \textit{point\_iq}\}
   \EndFor
   \For {\textit{point} in \textit{validPoints}}
   \For {iq in \textit{Path}}
   \If{ \textit{Evaluate}(iq, \textit{point}) = 0}\Comment{Point lies in the line}
   \State $I_p \leftarrow I_p \cup $ iq 
   \EndIf
   \EndFor
   \EndFor
   \State $I_c \leftarrow$ \textit{GetAllkCombinations}($I_p$, k)\Comment{Returns all k combinations from $I_p$}
   \For {each iq in $I_c$}
   \State $I_l \leftarrow $ \textit{GetAllSum}(iq) \hspace{0.2cm}\Comment{Form a new inequality after summing of k inequalities. It is also a linear combination of the original inequalities}
   
   \State \textit{point\_iq} $\leftarrow \phi$
   \For {\textit{point} in \textit{impPoints}}
   \If { \textit{Evaluate}($I_l$, \textit{point}) < 0}
   \State \textit{point\_iq} $\leftarrow$ \textit{point\_iq} $\cup$ \textit{point}
   \EndIf
   \EndFor
   \State \textit{Path} $\leftarrow$ \textit{Path} $\cup$ \{$I_l$, \textit{point\_iq}\}
   \EndFor
   
   \For { \textit{point} in \textit{impPoints} }
   \For { each \textit{path} in \textit{Path} } 
   \If{\textit{point} is in \textit{path}}
   \State \textit{coverMat[Path.inequality][Path.point]} $\leftarrow$ 1
   \EndIf
   \EndFor
   \EndFor
   \State Generate $m$ binary variables such that $c_1, c_2, \ldots , c_m$ such that $c_i = 1$ means inequality $i$ is included in the solution space else $c_i = 0$.
   \State \textit{constraintSet}$\leftarrow \phi$
    \For {$\textbf{each } point \in \textit{impPoints}$}
    \State \textit{constraintSet }$\leftarrow \textit{ constraintSet }  \cup  $ \{Construct a constraint $c_k,\ldots, c_l >= 1$ such that $p$ is removed by at least one inequality using matrix \textit{coverMat}\}\Comment{Generate constraints}
    \EndFor
    \State Create an objective function $\sum_{i=1}^m c_i$ with constraints \textit{constraintSet} and solve to get best inequalities ($I_{SR}$) that generate a stricter feasible region after maximizing the removed impossible differential patterns. 
    \State \textbf{return $I_{SR}$}
   \end{algorithmic}
   \caption{Subset Addition Algorithm}
   \label{algo:SubsetAddDetails}
\end{algorithm}

\section{Filtering Inequalities by Subset Addition}\label{SubsetAddition}
The main issue with the greedy algorithm is an optimal solution to minimisation problems is not guaranteed. Hence, we used Gurobi Optimizer~\cite{Gurobi}, as described by Sasaki and Todo~\cite{Sasaki_MILP_ALG}, to find the optimal solutions, and successfully replicated those results. Noting that solutions found in this manner are merely the optimal subsets of the H-representation, and not globally optimal, we attempt to
follow the technique proposed by Boura and Coggia~\cite{BouraC}. Their algorithm concentrates on producing a larger starting set of inequalities so that a smaller subset may more easily be found. They generates new inequalities by adding $k$-size subsets of inequalities. The new inequalities represents the hyperplanes which is the possible differential paths lie on. The newly generated inequalities are then discarded if they do not remove a new set of impossible differential paths. This is potentially slow, since it would necessitate comparing the lists of impossible differential paths removed by each of the constituent inequalities with that removed by the new inequality. Hence, we propose an alternative algorithm which differs from~\cite{BouraC} in the way new inequalities are assessed.
The concept is to add $k$ inequalities representing the hyperplanes a possible differential path lies on ($h_1$ through $h_k$), thus generating a new inequality
\begin{equation*}
    h=\sum_{i=1}^{k} h_i
\end{equation*}
and keep it only if it is good. We propose that h is good if,
\begin{itemize}
    \item \textbf{Type 1} New inequality $h$ removes more impossible differential paths than the inequality in ${h_1, h_2, h_3, \ldots, h_k}$ which removes the fewest; or
    \item \textbf{Type 2} New inequality $h$ invalidates at least as many impossible differential paths as the inequality in ${h_1, h_2, h_3, \ldots, h_k}$ which removes the most.
\end{itemize}

Algorithm~\ref{alg:subset_addition} explains the overall process. Finally, we find an optimal subset using Gurobi Optimizer. Note that, unlike~\cite{BouraC}, no regard is given to which impossible differential paths are removed by $h$. We only check how many of them it removes. Examples of the two types of hyperplanes are shown in Figure~\ref{fig:m_milp_diagram}. These are only illustrative examples, since differential paths lie on the vertices of the unit hypercube. $(h_1 + h_2$ is of type 2 as it
removes 12 impossible differential paths while $h_1$ removes removes 12, and $h_2$, 10; $h^{\prime}_1 + h^{\prime}_2$ is of type 1 as it removes 14 while $h^{\prime}_1$ and $h^{\prime}_2$ remove 16 and 10 respectively.

Now we describe our algorithm in details (see Algorithm~\ref{algo:SubsetAddDetails}). Assume \textit{ddt} be the 2D matrix for difference distribution table corresponding to the given SBox. Define $(u,v)$ as a point corresponding to the input difference ($u$) and output difference ($v$) in a \textit{ddt}. Let $(\Bar{u}, \Bar{v})$ be a valid transition and $(\Bar{\Bar{u}}, \Bar{\Bar{v}})$ be any impossible transition for an SBox. Divide the \textit{ddt} into two parts, $(\Bar{u}, \Bar{v})  \in$ \textit{validPoints} and $(\Bar{\Bar{u}}, \Bar{\Bar{v}}) \in$ \textit{impPoints}. Now apply \textit{validPoints} to the sagemath method \textit{inequality\_generator()}, which returns the inequalities $I$, corresponding to the H-representation of \textit{validPoints}. Assume a set \textit{Path}, where each element consists of two parts, an inequality $iq$ and a point $(u,v)$. 

Initially the set \textit{Path} is empty. Traverse each inequality $iq$ from set $I$. For each impossible point check if each inequality $iq$ satisfies or not. Assume point\_iq stores all the impossible points for an inequality $iq$. Finally, set \textit{Path} have all the inequalities and for each inequality the corresponding impossible points. Next, for each point in \textit{validPoints} perform a test if an inequality from set \textit{Path} satisfies the point. Collect all such inequalities in set $I_p$. Now use method \textit{GetAllkCombinations} to get all unique k-degree sets $I_c$. Next take elements from $I_c$ one by one and compute the sum of all the k inequalities. This will create a new inequality, which is linear combination of the original inequalities. Let the resultant inequality (sum) be $I_l$. Again we perform the same satisfiablity checking for the new inequality $I_l$ against all the impossible points from set \textit{impPoints}. Assume $I_l$ invalidates a subset point\_iq form the set \textit{impPoints}. We append $I_l$ with point\_iq to set \textit{Path}.
Now \textit{Path} contains the original inequalities as well as the newly generated inequalities along with the corresponding impossible differential points. Create a two-dimensional array \textit{coverMat} indicating which inequality removes which impossible differential paths. 

\begin{equation*}
    \textit{coverMat}_{i,j} = \begin{cases}
    1 \text{ if $i^{th}$ inequality removes $j^{th}$ impossible point}\\
    0 \text{ otherwise}
    \end{cases}
\end{equation*}

Derive an MILP problem to find the minimum subset of those inequalities which removes all impossible differential paths. We generate $m$ binary variables $c_1, c_2, \ldots , c_m$ such that $c_i = 1$ means inequality $i$ is included in the solution space else $c_i = 0$. For each point $p$ in \textit{impPoints} generate a constraint $c_k,\ldots, c_l >= 1$, such that $p$ is removed by at least one inequality using matrix \textit{coverMat} and store the constraints in \textit{constraintSet}.
Here the objective function is $\sum_{i=1}^m c_i$ with constraints \textit{constraintSet}. Now, solve the problem using an MILP tool to get the optimized set of inequalities $I_{SR}$, which generate a stricter feasible region after maximizing the removed impossible differential patterns.

\subsection{Multithreading and Filtration}
Each iteration of the loop (starting at line 4) of Algorithm \ref{alg:subset_addition} can run independently of any other. Consequently, the algorithm can be implemented in a multithreaded fashion, using a thread pool. Whenever any thread is free, it picks up the next available iteration of the loop and starts executing it. In doing so, we observed that one thread spends a noticeably longer time than the other threads, irrespective of the cipher under analysis. The reason is that the possible differential path corresponding to the origin: $[0, 0, 0, 0, 0, 0, 0, 0]$ appears to lie on significantly more hyperplanes than any other path; however, it does not generate any new inequalities which eventually form the optimal subset. As a result, the thread assigned to process it spends the longest amount of time doing nothing useful so that this path can be ignored from the outset.

\begin{table}
    \centering
    \caption{Minimum number of Inequalities for 4-bit SBoxes (Subset Addition)}
    \begin{tabular}{ |c|c|c|c|c| }
        \hline
        Cipher&\thead{Sasaki and\\ Todo~\cite{Sasaki_MILP_ALG}}&\thead{Boura and\\ Coggia~\cite{BouraC}}&\thead{Subset Addition ($k = 2$)\\(Our approach)}&\thead{Subset Addition ($k = 3$)\\(Our approach)}\\
        \hline
        GIFT&-&17&17&17\\
        KLEIN&21&19&19&19\\
        Lilliput&23&19&20&19\\
        MIBS&23&20&20&20\\
        Midori S0&21&16&17&16\\
        Midori S1&22&20&20&20\\
        Minalpher&22&19&19&\bf18\\
        Piccolo&21&16&16&16\\
        PRESENT&21&17&17&17\\
        PRIDE&-&16&17&17\\
        PRINCE&22&19&19&\bf18\\
        RECTANGLE&21&17&17&\bf16\\
        SKINNY&21&16&16&16\\
        TWINE&23&19&20&19\\
        \hline
        LBlock S0&24&17&17&\bf16\\
        LBlock S1&24&17&17&\bf16\\
        LBlock S2&24&17&17&\bf16\\
        LBlock S3&24&17&17&\bf16\\
        LBlock S4&24&17&17&\bf16\\
        LBlock S5&24&17&17&\bf16\\
        LBlock S6&24&17&17&\bf16\\
        LBlock S7&24&17&17&\bf16\\
        LBlock S8&24&17&17&\bf16\\
        LBlock S9&24&17&17&\bf16\\
        \hline
        Serpent S0&21&17&18&17\\
        Serpent S1&21&17&19&18\\
        Serpent S2&21&18&18&\bf17\\
        Serpent S3&27&20&16&\bf14\\
        Serpent S4&23&19&19&19\\
        Serpent S5&23&19&17&\bf17\\
        Serpent S6&21&17&16&\bf16\\
        Serpent S7&27&20&16&\bf16\\
        Serpent S8&-&-&18&18\\
        Serpent S9&-&-&18&17\\
        Serpent S10&-&-&17&16\\
        Serpent S11&-&-&15&15\\
        Serpent S12&-&-&18&18\\
        Serpent S13&-&-&18&18\\
        Serpent S14&-&-&18&18\\
        Serpent S15&-&-&18&18\\
        Serpent S16&-&-&17&16\\
        Serpent S17&-&-&19&19\\
        Serpent S18&-&-&18&18\\
        Serpent S19&-&-&18&17\\
        Serpent S20&-&-&19&19\\
        Serpent S21&-&-&18&17\\
        Serpent S22&-&-&17&16\\
        Serpent S23&-&-&19&19\\
        Serpent S24&-&-&18&17\\
        Serpent S25&-&-&17&16\\
        Serpent S26&-&-&18&18\\
        Serpent S27&-&-&17&16\\
        Serpent S28&-&-&17&17\\
        Serpent S29&-&-&17&17\\
        Serpent S30&-&-&17&17\\
        Serpent S31&-&-&18&17\\
        \hline
    \end{tabular}
    \label{tab:min_ineq_subsetaddition}
\end{table}

\subsection{Implementation and Results}
We implemented Algorithm~\ref{algo:SubsetAddDetails} in C++ (on a desktop computer with an Intel Core i5-6500 4C/4T
CPU running Manjaro Linux Sikaris 22.0.0 Xfce Edition 64-bit), and then extended the program
to use Gurobi Optimizer to find the optimal subset of inequalities. We give the user the option to
choose (by defining a macro while compiling) between good hyperplanes of types 1 and 2. Our
experiments suggest that selecting type 1 is better than or as good as selecting type 2.

\subsubsection*{Application to 4-bit SBoxes}
Algorithm~\ref{algo:SubsetAddDetails} is successfully applied to most of the 4-bit SBoxes. Among the 14 4-bit SBoxes provided in first block of Table~\ref{tab:min_ineq_subsetaddition}, we are getting better results for Prince~\cite{DBLP:prince}, Minalpher~\cite{Minalpher} and Rectangle~\cite{DBLP:rectangle} with setting $k=3$. For 10 SBoxes the results are same as the existing one. Only for PRIDE~\cite{PRIDE} the minimum number of inequalities is one extra.

For all the ten LBlock~\cite{DBLP:lblock} SBoxes the inequality count is decreased to 16 from 17~\cite{BouraC}. The results for LBlock are provided in block two in Table~\ref{tab:min_ineq_subsetaddition} for $k=2$ and $k=3$.

Block three of Table~\ref{tab:min_ineq_subsetaddition}) explains the results of 32 Serpent~\cite{DBLP:serpent} SBoxes. Comparing with the existing results of eight Serpent SBoxes $(S0 \text{ to }S7)$ we improve the results for five $(S2, S3, S5, S6, S7)$ For two SBoxes~$(S0, S4)$ the results are same, though for $S2$ we are loosing. For the rest of the 24 SBoxes we provide new results. For Serpent S3 the inequalities are mentioned in Appendix~\ref{Appendix_SampleInequalities}.

\subsubsection*{Application to 5- and 6-bit SBoxes}
We have applied Algorithm~\ref{algo:SubsetAddDetails} for ASCON~\cite{DBLP:ascon5} and SC2000~\cite{DBLP:sc2000-5}, which use 5-bit SBoxes. In this case, by taking $k=3$, the results are improved. For 6-bit SBoxes of APN~\cite{APN} and SC2000 with $k=2$, we can cross the existing boundary of Boura and Coggia~\cite{BouraC}. For 5-bit SBox of FIDES~\cite{FIDES} we are getting one extra inequality than the existing boundary. 
The results for 5 and 6-bit SBoxes are tabulated in Table~\ref{tab:min_ineq_5_6_bit_sbox} comparing with existing results. 

\subsubsection*{Reducing running time over Boura and Coggia~\cite{BouraC} technique}
As mentioned earlier, since each impossible differential path is processed independently, parallel processing can be
employed to reduce the running time. A worker thread can independently process a possible
differential path at a time.

\begin{table}[htbp]
\centering
\caption{ Minimum number of Inequalities for 5- and 6-bit SBoxes}
\label{tab:min_ineq_5_6_bit_sbox}
\begin{tabular}{|c|c|c|c|c|c|}
\hline
SBox   & \thead{SBox\\ Size} & SageMath~\cite{sagemath} & Boura and Coggia~\cite{BouraC} & \thead{Subset Addition\\ $k=2$\\(Our approach)} & \thead{Subset Addition\\ $k=3$\\(Our approach)} \\ \hline
ASCON  & \multirow{3}{*}{5}                                  & 2415     & 32           & \textbf{31}                                                   & \textbf{31}                                                    \\ \cline{1-1} \cline{3-6} 
FIDES  &                                                     & 910      & 61           & 64                                                   & 62                                                    \\ \cline{1-1} \cline{3-6} 
SC2000 &                                                     & 908      & 64           & 65                                                   & \textbf{63}                                                    \\ \hline
APN    & \multirow{3}{*}{6}                                  & 5481     & 167          & \textbf{163}                                                  & \multirow{3}{*}{--}                                   \\ \cline{1-1} \cline{3-5}
FIDES  &                                                     & 7403     & 180          & 184                                                  &                                                       \\ \cline{1-1} \cline{3-5}
SC2000 &                                                     & 11920    & 214          & \textbf{189}                                                  &                                                       \\ \hline
\end{tabular}
\end{table}

For multithreading, we used the C++ POSIX Threads API (which is wrapped in the thread
library). Boura and Coggia~\cite{BouraC} reported that for $k = 2$, their algorithm implementation took a few minutes, while for $k = 3$, it took a few hours. Having made all optimisations mentioned above in our program, we achieved much better running times. In Table~\ref{table:m_milp_time} we have tabulated the running time of our algorithm (Algorithm ~\ref{algo:SubsetAddDetails}) for some Sboxes. For LBlock S0 through S9 and Serpent S0 through S7, average running times are reported. In general, it appears that larger values of $k$ lead to smaller subsets of inequalities. However,
we were unable to confirm this. While $k = 3$ usually produced smaller subsets than $k = 2$ in our
experiments, testing with $k = 4$ proved difficult. For Lilliput~\cite{Lilliput}, MIBS~\cite{DBLP:mibs} and Serpent S3, the outputs
did not improve with $k = 4$, but the memory requirement shot up to around 10 GiB. We could not
test any other ciphers because of this. We compare our algorithm for $k = 3$ with that of Boura and Coggia~\cite{BouraC} in Table~\ref{table:m_milp_time}. Our implementation is faster by two orders of magnitude, and gives comparable results.

\begin{table}[htbp]
\centering
\caption{Approximate running time of Subset Addition Algorithm}
\label{table:m_milp_time}
\begin{tabular}{|c|cc|}
\hline
\multirow{2}{*}{SBox} & \multicolumn{2}{c|}{Required Time for Algorithm~\ref{algo:SubsetAddDetails} (in sec)} \\ \cline{2-3} 
                      & \multicolumn{1}{c|}{No. of Inequality, k = 2} & No. of Inequality, k = 3                          \\ \hline
Klein                 & \multicolumn{1}{c|}{0.16}                           & 2.5                          \\ \hline
LBlock S*             & \multicolumn{1}{c|}{0.19}                           & 2.2                          \\ \hline
MIBS                  & \multicolumn{1}{c|}{1.9}                            & 4.5                          \\ \hline
Piccolo               & \multicolumn{1}{c|}{0.15}                           & 2.0                          \\ \hline
PRESENT               & \multicolumn{1}{c|}{0.28}                           & 3.9                          \\ \hline
PRINCE                & \multicolumn{1}{c|}{0.17}                           & 4.8                          \\ \hline
Serpent S*            & \multicolumn{1}{c|}{0.49}                           & 8.3                          \\ \hline
TWINE                 & \multicolumn{1}{c|}{0.16}                           & 3.4                          \\ \hline
\end{tabular}
\end{table}

\section{Conclusion}\label{Conclusion}
In this paper, we propose two MILP-based techniques for finding the minimum set of inequalities for modeling differential propagations of an SBox. The algorithms we introduce for modeling the DDT of an SBox are more efficient than the other existing algorithms. Noting that a greedy algorithm is only complete with the notion of a tiebreaker, we implemented a new version of the greedy approach based on a random tiebreaker. The results of the greedy random tiebreaker outperform the original greedy one for some of the SBoxes. The subset addition algorithm can successfully model SBoxes up to 6-bit. The approach also provides better or almost identical results for most SBoxes. We also improved the execution time to find the minimized inequalities concerning the previous implementations.  

%

\bibliographystyle{splncs04}
\bibliography{main}

\appendix
\section{Existing Algorithm for Choosing Best Inequalities}\label{Appendix_Algorithms}

\begin{algorithm}[H]
    Input:\\ 
    \textit{HI}: Inequalities in the H-Representation of the convex hull of an SBox.\\
    \textit{ID}: The set of all impossible differential patterns of an SBox 
    \\
    Output:\\
    $RI$: Set of inequalities that generates a stricter feasible region after maximizing the removed impossible differential patterns. 
	\begin{algorithmic}[1]
   \State $l \gets \phi, \textit{RI} \gets \phi$
    \While {\textit{ID} $\ne \phi$}
    \State $l \gets $ The inequality in \textit{HI} which maximizes the number of removed impossible differential patterns from \textit{ID}.
    \State \textit{ID} $\gets$ \textit{ID} $-$ \{\text{Removed impossible differential patterns using $l$\}}
    \State \textit{HI} $\gets$ \textit{HI} $-$ \{l\}
    \State \textit{RI} $\gets$ \textit{RI} $\cup$ \{l\}
    \EndWhile
    \State \textbf{return} \textit{RI}
    \end{algorithmic}
    \caption{Greedy Based Approach~\cite{Sun_asiacrypt}}
	\label{alg:greedy_sun}
	\end{algorithm}

\begin{algorithm}[H]
    Inputs:\\ 
    \textit{IDP}: Impossible differential patterns corresponding to the impossible transitions from the DDT of an SBox.\\
    $P$: Input set corresponding to the possible transitions in an SBox
    \\
    Output:\\
    $I_{SR}$: Set of inequalities that generates more stricter feasible region after maximizing the removed impossible differential patterns. 
	\begin{algorithmic}[1]
    \State H $\gets \textit{ConvHull(P)}$
    \State \textit{RI} $\gets H$
    \State Create a table \textit{PIT} of size  $|\textit{RI}| \times |\textit{IDP}|$ where $\textit{PIT}_{i,j} = 1$ if inequality \textit{$RI_i$} removes pattern $\textit{IDP}_j$, else set $\textit{PIT}_{i,j} = 0$
    \State Set $m= |\textit{RI}|$ 
    \State Generate $m$ binary variables such that $c_1, c_2, \ldots , c_m$ such that $c_i = 1$ means inequality $i$ is included in the solution space else $c_i = 0$.
   \State \textit{constraintSet}$\leftarrow \phi$
    \For {$\textbf{each } point \in \textit{IDP}$}
    \State \textit{constraintSet }$\leftarrow \textit{ constraintSet }  \cup  $ \{Construct a constraint $c_k,\ldots, c_l >= 1$ such that $p$ is removed by at least one inequality applying table \textit{PIT}\} \Comment{Generate constraints}
    \EndFor
    \State Create an objective function $\sum_{i=1}^m c_i$ with constraints \textit{constraintSet} and solve to get best inequalities ($I_{SR}$) that generate a stricter feasible region after maximizing the removed impossible differential patterns. 
    \State \textbf{return $I_{SR}$}
    \end{algorithmic}
    \caption{MILP based reduction~\cite{Sasaki_MILP_ALG}}
	\label{alg:sasaki_MILP}
	\end{algorithm}

 \begin{algorithm}[H]
    Inputs:\\ 
    $P$: Input set corresponding to the possible transitions in an SBox.\\
    $k$: The number of inequalities to be added together.
    \\
    Output:\\
    $RI$: Set of inequalities that generates more stricter feasible region after maximizing the removed impossible differential patterns. 
	\begin{algorithmic}[1]
    \State $H \gets \textit{ConvHull}(P)$
    \State \textit{RI} $\gets H$
    \For {$\textbf{all } p \in P$}
    \State Choose $k$ inequalities such that $p$ belongs to the hyperplanes of $Q_1,Q_2,\ldots,Q_k$
    \State $Q_{new}=Q_1+\ldots+Q_k$
    \State \textbf{if } $Q_{new}$ removes new impossible transitions
    \State \textit{RI} $\gets$ \textit{RI} $\cup\{Q_{new}\}$
    \State \textbf{end if}
    \EndFor
    \State \textbf{return \textit{RI}}
    \end{algorithmic}
    \caption{Modeling by selecting random set of inequalities~\cite{BouraC}}
	\label{alg:boura_select_any}
	\end{algorithm}

\section{Sample Reduced Inequalities}\label{Appendix_SampleInequalities}
Applying random greedy tiebreaker algorithm~\ref{alg:greedy_random} for MIBS~\cite{DBLP:mibs}, the reduced 24 inequalities are as follows,
\begin{verbatim}
 - 1x3 - 2x2 - 2x1 - 1x0 + 4y3 + 5y2 + 5y1 + 5y0 >= 0
 + 5x3 + 4x2 + 4x1 + 3x0 - 1y3 - 2y2 + 1y1 - 2y0 >= 0
 - 2x3 + 2x2 + 4x1 + 1x0 + 3y3 + 1y2 - 3y1 - 3y0 >= -4
 - 1x3 - 4x2 + 3x1 + 2x0 - 1y3 - 3y2 + 4y1 + 2y0 >= -5
 - 2x3 + 1x2 - 3x1 - 1x0 - 1y3 - 3y2 - 2y1 - 2y0 >= -11
 - 1x3 - 2x2 - 4x1 + 4x0 - 4y3 + 2y2 + 1y1 - 3y0 >= -10
 + 2x3 - 1x2 + 3x1 + 1x0 - 2y3 + 2y2 - 3y1 + 1y0 >= -3
 + 1x3 + 2x2 - 4x1 + 2x0 + 3y3 + 1y2 + 2y1 + 4y0 >= 0
 + 1x3 + 3x2 - 2x1 - 3x0 + 1y3 + 3y2 + 2y1 - 1y0 >= -3
 + 2x3 - 1x2 - 2x1 - 2x0 - 1y3 - 1y2 - 2y1 + 0y0 >= -7
 + 0x3 + 2x2 + 2x1 - 1x0 + 1y3 + 1y2 - 1y1 + 1y0 >= 0
 - 3x3 - 3x2 + 1x1 - 2x0 + 1y3 - 2y2 + 1y1 + 2y0 >= -7
 + 2x3 - 1x2 + 2x1 - 1x0 + 1y3 + 1y2 + 2y1 - 1y0 >= -1
 + 1x3 - 2x2 - 2x1 + 2x0 + 1y3 + 1y2 - 1y1 - 2y0 >= -5
 - 1x3 + 2x2 - 1x1 + 1x0 + 2y3 - 2y2 + 1y1 - 1y0 >= -3
 - 1x3 + 1x2 + 0x1 - 1x0 - 1y3 - 1y2 + 0y1 + 1y0 >= -3
 + 1x3 - 2x2 - 1x1 - 1x0 + 1y3 - 2y2 - 2y1 + 1y0 >= -6
 + 2x3 - 1x2 + 0x1 - 2x0 - 2y3 + 2y2 - 1y1 + 1y0 >= -4
 - 1x3 - 1x2 + 1x1 - 1x0 - 1y3 + 0y2 - 1y1 - 1y0 >= -5
 - 1x3 + 1x2 - 1x1 + 2x0 + 1y3 + 2y2 - 1y1 + 2y0 >= -1
 + 2x3 + 1x2 + 2x1 + 3x0 - 2y3 - 1y2 - 1y1 + 2y0 >= -1
 - 3x3 - 2x2 + 1x1 + 3x0 - 1y3 + 1y2 + 2y1 + 3y0 >= -3
 + 1x3 - 1x2 - 2x1 - 2x0 - 1y3 - 1y2 - 1y1 - 1y0 >= -7
 - 1x3 + 1x2 + 0x1 - 1x0 - 1y3 + 1y2 + 1y1 - 1y0 >= -3
 \end{verbatim}
Applying subset addition algorithm~\ref{algo:SubsetAddDetails} for Serpent S3 the inequalities are as follows,
\begin{verbatim}
 - 5x3 + 4x2 + 4x1 - 5x0 + 2y3 + 10y2 + 3y1 + 10y0 >= 0
 + 6x3 - 1x2 - 2x1 + 2x0 + 1y3 +  7y2 - 3y1 +  7y0 >= 0
 - 2x3 + 0x2 - 3x1 - 3x0 - 2y3 -  4y2 - 1y1 +  4y0 >= -11
 + 3x3 + 0x2 + 3x1 + 2x0 + 1y3 -  4y2 + 2y1 +  4y0 >= 0
 - 3x3 - 3x2 + 0x1 - 2x0 - 1y3 +  4y2 - 2y1 -  4y0 >= -11
 - 4x3 - 4x2 - 1x1 - 3x0 + 1y3 +  2y2 - 1y1 -  4y0 >= -13
 + 2x3 - 2x2 + 1x1 - 4x0 - 4y3 +  3y2 + 2y1 -  4y0 >= -10
 + 2x3 + 6x2 + 2x1 + 1x0 - 3y3 -  4y2 - 4y1 -  4y0 >= -10
 - 2x3 + 8x2 + 4x1 - 1x0 + 5y3 -  7y2 + 6y1 -  7y0 >= -10
 - 2x3 - 5x2 - 1x1 + 2x0 - 3y3 -  5y2 + 3y1 -  5y0 >= -17
 + 2x3 + 3x2 + 0x1 + 3x0 + 2y3 +  4y2 + 1y1 -  4y0 >= 0
 + 4x3 - 3x2 - 2x1 + 0x0 + 2y3 -  3y2 - 1y1 -  3y0 >= -9
 - 2x3 - 1x2 + 2x1 + 4x0 + 4y3 -  4y2 - 2y1 +  3y0 >= -5
 + 0x3 - 1x2 - 1x1 + 5x0 - 2y3 +  5y2 + 2y1 +  5y0 >= 0
 \end{verbatim}

%
%

\end{document}